\def\expec#1{\langle#1\rangle}

\def\bigexpec#1{\left\langle#1\right\rangle}

\def\eg{{\frenchspacing\it e.g.}}
\def\etc{{\frenchspacing\it etc.}}

\def\rf#1;#2;#3;#4;#5 {\par#1, {#3} {\bf #4}, #5 (#2). \par}

\def\beq#1{\begin{equation}\label{#1}}
\def\eeq{\end{equation}}
\def\beqa#1{\begin{eqnarray}\label{#1}}
\def\eeqa{\end{eqnarray}}
\def\eq#1{{\frenchspacing eq.}~(\ref{#1})}
\def\Eq#1{{\frenchspacing Eq.}~(\ref{#1})}
\def\eqno#1{~(\ref{#1})}

\def\C{{\bf C}}
\def\F{{\bf F}}
\def\Veff{V_{eff}}
\def\nbar{{\bar n}}
\def\r{{\bf r}}
\def\x{{\bf x}}
\def\kmin{{k_{min}}}
\def\kmax{{k_{max}}}
\def\tr{\hbox{tr}\>}
\font\bfmath=cmmib10
\def\vth{
\hbox{\bfmath\char'002}}  	
\def\vmu{
\hbox{\bfmath\char'026}}	
\def\Mpc{{\rm Mpc}}

\documentstyle[prl,aps,epsf]{revtex}
\begin{document}
\twocolumn[\hsize\textwidth\columnwidth\hsize\csname@twocolumnfalse\endcsname


\preprint{IASSNS-AST 97/666}

\title{Measuring Cosmological Parameters with Galaxy Surveys}

\author{Max Tegmark}

\address{Institute for Advanced Study, Princeton, NJ 08540; max@ias.edu}

\date{Submitted 18 June 1997; published 17 November}

\maketitle

\begin{abstract}
We assess the accuracy with which future galaxy surveys can
measure cosmological parameters. 
By breaking parameter degeneracies
of the Planck Cosmic Microwave Background satellite, 
the Sloan Digital Sky Survey may be able to reduce the Planck error bars 
by about an order of magnitude on the large-scale power
normalization and the reionization optical depth,
down to percent levels. 
However, pinpointing attainable 
accuracies to within better than a factor of a few
depends crucially on whether it will be possible to 
extract useful information from the mildly nonlinear regime.

\end{abstract}


] 

\makeatletter
\global\@specialpagefalse
\def\@oddfoot{
\ifnum\c@page>1
  \reset@font\rm\hfill \thepage\hfill
\fi
\ifnum\c@page=1
Published in {\it\frenchspacing Phys. Rev. Lett.} {\bf 79}, 3806 (1997).
Available in color from 
{\it h t t p://www.sns.ias.edu/$\tilde{~}$max/galfisher.html} \hfill\\
\fi
} \let\@evenfoot\@oddfoot
\makeatother

\section{INTRODUCTION}

One of the main challenges in modern cosmology is to refine and test the standard 
gravitational instability model of structure formation by precision measurements
of its free parameters: the slope $n$ and normalization $Q$ 
of the primordial spectrum of density fluctuations, the densities of various
types of matter, {\etc}
A seminal paper 
in this journal 
\cite{Jungman96a} recently showed that
future cosmic microwave background (CMB) 
experiments such as the MAP and Planck satellites
would revolutionize this endeavor, allowing the
simultaneous determination of a dozen parameters to hitherto unprecedented 
accuracies. This prompted several more detailed studies 
\cite{Jungman96b,parameters,Zaldarriaga97}, which confirmed this
optimistic conclusion.

A parallel effort towards precision cosmology is larger and more
systematic galaxy redshift surveys. The largest currently available 
three-dimensional surveys
contain about 25,000 galaxies. The 2dF survey
(described in \cite{Strauss97}) will measure ten times as many,
and the Sloan Digital Sky Survey (SDSS) is scheduled to acquire a million 
redshifts within five years \cite{Strauss97,Gunn95}. 
It is therefore quite timely to perform an analogous   
first assessment of the ability to measure cosmological 
parameters with large galaxy surveys. 
This is the purpose of the present {\it Letter}. 

\section{METHOD}

The accuracy with which cosmological parameters can be measured 
from a given data set is conveniently computed with the 
Fisher information matrix formalism (see \cite{karhunen} for a comprehensive review).
In our case, the data set can be viewed as an $N$-dimensional
vector $\x$, whose components $x_i$ are the 
fluctuations in the galaxy density relative to the mean 
in $N$ disjoint cells that cover the three-dimensional 
survey volume in a fine grid. $\x$ is modeled as a random variable whose probability
distribution $f(\x;\vth)$ depends on a vector of cosmological parameters $\vth$
that we wish to estimate (for instance, we might have $\theta_1=n$, 
$\theta_2=Q$, \etc).
The Fisher matrix is defined by 
\beq{FisherDefEq}
\F_{ij} \equiv - \bigexpec{{\partial^2 \ln f\over\partial\theta_i\partial\theta_j}},
\eeq
and its inverse $\F^{-1}$ 
can, crudely speaking, be thought of as the
best possible covariance matrix for the measurement errors on the parameters.
The Cram\'er-Rao inequality 
\cite{KendallStuart69}
shows that no unbiased method whatsoever
can measure the $i^{th}$ parameter with error bars (standard
deviation) less than $1/\sqrt{\F_{ii}}$.
If the other parameters are not known but are estimated
from the data as well, the minimum standard deviation
rises to $(\F^{-1})_{ii}^{1/2}$.

\subsection{The brute force approach}

In the approximation that the probability distribution $f$ is a multivariate Gaussian
with mean $\vmu\equiv\expec{\x}$ and covariance matrix
$\C\equiv\expec{\x\x^t}-\vmu\vmu^t$, 
\eq{FisherDefEq}
becomes \cite{VS96,karhunen}
\beq{GaussFisherEq}
\F_{ij} = {1\over 2}\tr
\left[\C^{-1}{\partial\C\over\partial\theta_i}\C^{-1}{\partial\C\over\partial\theta_j}\right]
+ {\partial\vmu\over\partial\theta_i}^t\C^{-1}{\partial\vmu\over\partial\theta_j}.
\eeq
This equation was employed in all the above-mentioned papers
on CMB parameter determination, since for an all-sky CMB map, 
the covariance matrix $\C$ can be diagonalized by a spherical
harmonic expansion, making the computation of $\F$ numerically trivial. 
For our galaxy survey case, the situation is more difficult.
The analog of the CMB trick (a Fourier transformation) does not diagonalize
$\C$, since only a finite spatial volume is probed.

\subsection{A useful approximation}

Since brute force application of \eq{GaussFisherEq}
tends to obscure the underlying physics, we will now derive a simple approximation
for $\F$ below, which allows a more intuitive understanding of numerical results
and shows the relative information contribution from 
different scales $k$.
Ignoring redshift-space distortions and non-linear 
clustering, all the cosmological information is contained in
the galaxy power spectrum $P(k)$.
In the limit where the survey volume is much larger than the scale of 
any features in $P(k)$, it has been shown 
\cite{Hamilton97a} 
that all the cosmological information in $\x$ is recovered when $P(k)$ is 
estimated with the FKP method \cite{FKP}.
Let us therefore redefine $x_n$ to be not the density fluctuation
in the $n^{th}$ spatial volume element, but the
average power measured with the FKP method in a thin shell 
of radius $k_n$ in Fourier space, with width $dk_n$ and volume element 
$V_n\equiv 4\pi k_n^2 dk_n/(2\pi)^3$. With our notation, we can rewrite the FKP
results \cite{FKP} as
\beqa{FKPmeanEq}
\vmu_n&\approx&P(k_n),\\
\label{FKPcovEq}
\C_{mn}&\approx&2{P(k_n) P(k_n)\over V_n\Veff(k_n)}\,\delta_{mn},
\eeqa
where 
\beq{VeffDefEq}
\Veff(k)\equiv\int\left[{\nbar(\r)P(k)\over 1+\nbar(\r)P(k)}\right]^2 d^3r.
\eeq
Here $\nbar(\r)$ is the selection function of the survey, 
which gives the {\it a priori} expectation value for the number 
density of galaxies. $\Veff(k)$ can be interpreted as the effective
volume utilized for measuring the power at wavenumber $k$, since
the integrand will be of order unity in those regions where
the cosmic signal $P(k)$ exceeds the Poissonian shot noise
$1/\nbar$, and typically gives only a small contribution 
from other regions.
For a volume-limited survey, $\nbar$ is constant in the observed region, so 
$\Veff$ (and hence the Fisher matrix) is simply proportional
to the survey volume.

Choosing the shells thick enough to contain many uncorrelated modes each,
$V_n V_{eff}(k_n) \gg 1$,
the central limit theorem indicates that $\x$ will be approximately Gaussian.
In the same limit $V_n V_{eff}(k_n)\gg 1$,
the second term in \eq{GaussFisherEq}
will be completely dominated by the first \cite{cl}, so substituting
equations\eqno{FKPmeanEq} and\eqno{FKPcovEq} into \eq{GaussFisherEq} gives
\beq{FisherDerivationEq}
\F_{ij} \approx  {1\over 4\pi^2}\sum_n 
{\partial P\over\partial\theta_i}(k_n) 
{\partial P\over\partial\theta_j}(k_n)
{\Veff(k_n)k_n^2 dk_n\over P(k_n)^2}. 
\eeq
Replacing the sum by an integral and using 
$d\ln P=dP/P$, 
this reduces to the handy approximation
\beq{PowerRangerEq}
\F_{ij}\approx 2\pi\int_\kmin^\kmax
{\partial\ln P\over\partial\theta_i}{\partial\ln P\over\partial\theta_j}
w(k)d\ln k,
\eeq
where we have defined 
\beq{wDefEq}
w(k)\equiv {\Veff(k)\over\lambda^3},
\eeq
and the wavelength is $\lambda\equiv 2\pi/k$.
\Eq{PowerRangerEq} conveniently separates the 
effects of cosmology from those of the survey-specific details.
The former enter only through the logarithmic
derivatives $\partial\ln P/\partial\theta_i$, which are 
plotted in Figure 1 for some simple examples.
The selection function $\nbar$ and 
the geometric bounds of the survey volume (outside of
which $\nbar=0$) enter only via the weight function $w(k)$,
which is essentially the number of independent modes of wavelength
$\lambda$ that fit into the volume probed ($\Veff$). 
The top panel of Figure 1 shows the weight function for the
main northern part of the SDSS \cite{Gunn95} and for the SDSS bright
red galaxy (BRG) sample. The latter is assumed to be 
volume-limited at 1000 $h^{-1}\Mpc$, containing
$10^5$ galaxies with a bias factor $b=2$, 


We close this section by emphasizing that \eq{PowerRangerEq} 
is a rather crude approximation, since it ignores edge effects, 
redshift space distortions and, most importantly, non-linear clustering. 
We will return to the last issue in Section~\ref{NonlinSec}.
To quantify the edge effect errors, 
we have tested \eq{PowerRangerEq}
numerically by brute force manipulations of the $N\times N$ matrices
of \eq{GaussFisherEq} for a number of cases with $N\sim 10^4$, 
and find that it is typically 
accurate to within a factor of two for a 
cold dark matter (CDM) power spectrum when the survey size
$\gg  200 h^{-1}$ Mpc.
The differences have two sources, with opposite sign, which
both grow in importance if we decrease the survey volume:
\begin{enumerate}
\item 
The effective number of modes probed is slightly larger
than $\Veff$ indicates, since the density field just inside the survey
volume is correlated with that just outside. This reduces error bars.
\item
The measured power spectrum is effectively smoothed on the scale
of the survey volume, which can destroy 
information on the small $k$ behavior of the power spectrum and on 
sharp features and wiggles. This increases error bars.
\end{enumerate}

\section{RESULTS AND CONCLUSIONS}

\subsection{A linear clustering example}

Before discussing realistic non-linear power spectra,
we will now highlight some of the features of 
\eq{PowerRangerEq} with a simple linear power spectrum example.
Let us consider a CDM power spectrum of the form 
\beq{LinearModelEq}
P(k) = Q^2 (\eta k/k_*)^n T(\eta k)^2.
\eeq
On a log-log plot such as Figure 1 (top panel), varying 
the normalization $Q$ shifts the spectrum vertically, whereas  
varying the parameter $\eta$ shifts it horizontally.
We chose $k_*=0.025h\Mpc^{-1}$, roughly the scale where 
$P$ takes its maximum, so varying $n$ tilts the spectrum about its peak.
The transfer function $T$ is computed numerically
with the CMBFAST software \cite{cmbfast} for a 
Hubble constant $h=0.5$, baryon fraction $\Omega_b=0.06$, 
CDM fraction $\Omega_c=0.48$, and vacuum density 
(relative cosmological constant) $\Omega_v=0.46$, chosen to be
virtually indistinguishable from a Bond \& Efstathiou model fit
\cite{BE84} with ``shape parameter'' $\Gamma\equiv h\Omega_c=0.25$.
For our fiducial model, $n=\eta=1$ and $Q$ is such that
the $8h^{-1}\Mpc$ normalization is $\sigma_8=1$.

Partial derivatives needed for \eq{PowerRangerEq} are plotted in the 
second panel of Figure 1.
$\partial{\ln P/\partial \ln Q}=2,$ 
$\partial{\ln P/\partial n}=\ln(k/k_*)$, and
$\partial{\ln P/\partial\ln\eta}=\partial{\ln P/\partial\ln k}$,
simply the logarithmic slope of the power spectrum, ranging from 
$+1$ to $-3$ and vanishing at the peak (together with $\partial{\ln P/\partial n}$).
The dependence on all other parameters $\theta_i$ enters via the transfer
function. Figure 1 shows only one such example: the baryon fraction $\Omega_b$.

\subsubsection{Single-parameter accuracy}

The third panel in Figure 1 shows the error bars 
$\Delta\theta_i=1/\F_{ii}^{1/2}$ on each parameter that would  
result from the SDSS BRG survey 
if the true values of all other parameters where known,
as a function of the upper limit of integration $\kmax$, with 
$\kmin=0$. As \eq{PowerRangerEq} shows, the information $\F_{ii}$ on 
a parameter is simply the square of the corresponding curve in 
the second panel, integrated against the weight function in the 
top panel. For instance, there is no information about $\Omega_b$ on
scales $k\ll k_*$, since the physical impact of baryons on 
fluctuation growth is different
from that of CDM only on scales entering the horizon before 
matter and radiation decouple at $z\sim 10^3$ \cite{HS96}.
Also, we see that the bulk of the information on $\Omega_b$ is
coming not from the characteristic baryon-induced 
acoustic oscillations (wiggles) in the transfer function, 
but from the overall suppression of power rightward of the peak. 
Although the wiggles help somewhat in breaking parameter degeneracy 
(discussed below), this can be somewhat misleading, since 
all but perhaps the first oscillation are
likely to have been smeared out by mode coupling as 
the clustering goes nonlinear. 
For a more detailed treatment of
the constraints on $\Omega_b$, submitted after the present paper,
see \cite{Goldberg97}.

How should the limits of integration ($\kmin$ and $\kmax$) be chosen?
Since information on scales comparable to and larger than the survey is 
destroyed by smearing and mean removal effects,
it is natural to chose $2\pi/\kmin$ to be of order
the survey size. 
The choice of $\kmax$, on the other hand, is seen to be of paramount importance, 
since the $k^3$ phase space factor causes $w(k)$ to peak far
shortward of the power spectrum peak scale $k_*$, where nonlinear effects 
become important. We defer this issue to Section~\ref{NonlinSec}.

\subsubsection{Degeneracies}

The bottom panel in Figure 1 shows the error bars 
$\Delta\theta_i=(\F^{-1})_{ii}^{1/2}$ on each parameter that would  
result if a joint fit to all four parameters were performed, 
and no other constraints (\eg, from CMB maps) 
were available for the other three parameters.
\Eq{PowerRangerEq} can be interpreted as $\F$ being the dot products
of a set of vectors (the functions $\partial\ln P/\partial\theta_i$),
where the inner product is defined by the weight function $w$.
If any of the functions in the second panel can be written as a linear
combination of some others, then $\F$ will clearly be singular, and
the errors on the corresponding parameters will
be infinite. For instance, 
$\partial\ln P/\partial\ln\eta$ and $\partial\ln P/\partial n$ 
are essentially degenerate for $\kmax<0.1 h\Mpc^{-1}$
(they both look like straight lines vanishing at
$k=k_*$, and the curvature of $\partial\ln P/\partial\ln\eta$
at $k<0.01 h\Mpc^{-1}$ is irrelevant since these scales receive
so little weight), which is why $\ln\eta$ and $n$ have such large 
uncertainties in the bottom panel until $\partial\ln P/\partial\ln\eta$
bends downward and breaks this near degeneracy at $k\sim 0.1 h\Mpc^{-1}$.

\subsection{Non-linear clustering}

\label{NonlinSec}

Since much of the information on cosmological parameters comes from 
small scales, non-linear clustering cannot be ignored
when assessing the attainable accuracy.
The power spectrum remains a perfectly well-defined  
quantity even in the deeply non-linear regime. However, 
the density field becomes non-Gaussian, which causes 
\eq{GaussFisherEq} (and hence also \eq{PowerRangerEq})
to misestimate the Fisher matrix in two competing ways:
\begin{enumerate}
\item The variance of the power spectrum estimates tend to exceed the value 
given by \eq{FKPcovEq}, causing us to underestimate the parameter error bars.
\item Additional cosmological information is contained in the higher
moments of the distribution, causing us to overestimate the parameter error bars.
\end{enumerate}
Bearing these important caveats in mind, we nonetheless apply \eq{PowerRangerEq},
using the analytic fits described in \cite{Jain95} to compute the relevant 
nonlinear power spectra. This changes the accuracy curves corresponding to Figure 1 
for $k\gg 0.1h\Mpc^{-1}$, but only marginally.
A more radical change occurs when including the linear bias factor $b$ 
(the ratio of the galaxy fluctuations to the underlying matter fluctuations, which we 
assume to be scale-independent) as an additional parameter, 
since in linear theory, it is perfectly degenerate with the 
large-scale power normalization $Q$.
The top panel of Figure 2 shows the partial power derivatives with respect to 
$Q$, $b$, $n$ and $\eta$, and we see that nonlinear effects begin
to break this degeneracy around the scale $k\sim 0.1h\Mpc^{-1}$.
Power spectra with wiggles cannot be accurately treated with this
nonlinear formalism, so we have 
used the above-mentioned wiggle-free 
Bond \& Efstathiou transfer function fit \cite{BE84}
here to be conservative and avoid underestimating error bars.


\subsection{Combining galaxy surveys and CMB experiments}

So what is the bottom 
line? How well can future galaxy surveys constrain 
cosmological parameters? Since degeneracies are crucial, especially 
when considering joint fits to a dozen parameters
as in the context of CMB experiments, a sensible answer must clearly take
into account the degeneracy-breaking information from other sources. 
It has recently been shown \cite{parameters,Zaldarriaga97} that CMB experiments 
suffer from a near-exact degeneracy between the spatial curvature
$\Omega$ and the cosmological constant $\Lambda$ (since they are virtually unable 
to distinguish between combinations that give the same angle-distance relationship),
but this degeneracy is likely to be independently broken by both
supernova and lensing measurements. The second worst degeneracy for the Planck
satellite links the normalization $Q$ to $\tau$ 
(the optical depth from reionization), and partly also to the scalar-to-tensor 
ratio. This second degeneracy is an example where future galaxy surveys have the potential
to substantially improve the situation. The bottom panel of Figure 2 shows 
the error bars on $b$ and $Q$ when $n$ and $\eta$ are assumed known, 
and it is seen that an accuracy $\Delta Q/Q=1\%$ is 
attained for 
$\lambda=2\pi/k\sim 18h^{-1}\Mpc$.
A fundamental limit on $Q$-accuracy will probably arise from partial 
degeneracy with the location and slope of the spectrum ($\eta$ and $n$)
on small scales, so since these parameters
can only be measured to about $1\%$ by Planck \cite{parameters},  
the $Q$-accuracy from SDSS will at best be of the same order.
Hovever, if this accuracy is indeed attainable despite the above-mentioned
caveats regarding nonlinearity, it would be quite a radical improvement 
over the $\Delta Q/Q\sim 15\%$ that Planck alone  
can attain \cite{parameters}.
By breaking this degeneracy, SDSS would also help Planck pin
down the other parameters that were nearly degenerate with $Q$.
For instance, repeating the analysis of \cite{parameters} with a 
mere $1\%$ prior uncertainly on $Q$, we find that the error bar
on the reionization optical depth drops from 0.16 to 0.03,
which would make reionization detectable at $1-\sigma$
as late as $z=8$ in a standard CDM cosmology.

In conclusion, we have derived, tested and applied an approximate formula
for the accuracy with which large galaxy surveys can measure cosmological
parameters. Although our results indicate that such
surveys can substantially enhance the accuracy attainable 
from CMB measurements alone, a number of issues must be addressed
before 
quantitative claims should be believed.
\begin{enumerate}
\item Are current calculations \cite{Jain95} of the non-linear power
spectrum sufficiently accurate for our application
(when including the effect of baryons, possible massive 
neutrinos, {\etc})?
\item Does the non-Gaussianity of the cosmological density field on 
weakly nonlinear scales cause our approximation to substantially over- or 
under-estimate the attainable accuracy?
\item Is biasing sufficiently non-linear \cite{Weinberg95} on these scales to  
invalidate our results?
\end{enumerate}

Thus although \eq{PowerRangerEq} is in itself a rather crude approximation, 
the main source of uncertainty lies elsewhere: in our ability to model 
and extract information from clustering in the marginally non-linear regime.
The nonlinear domain appears to be a gold mine of cosmological information,
but one whose riches may prove extremely difficult to extract.



\clearpage
\onecolumn
\begin{figure}[phbt]
\centerline{{\vbox{\epsfxsize=20cm\epsfbox{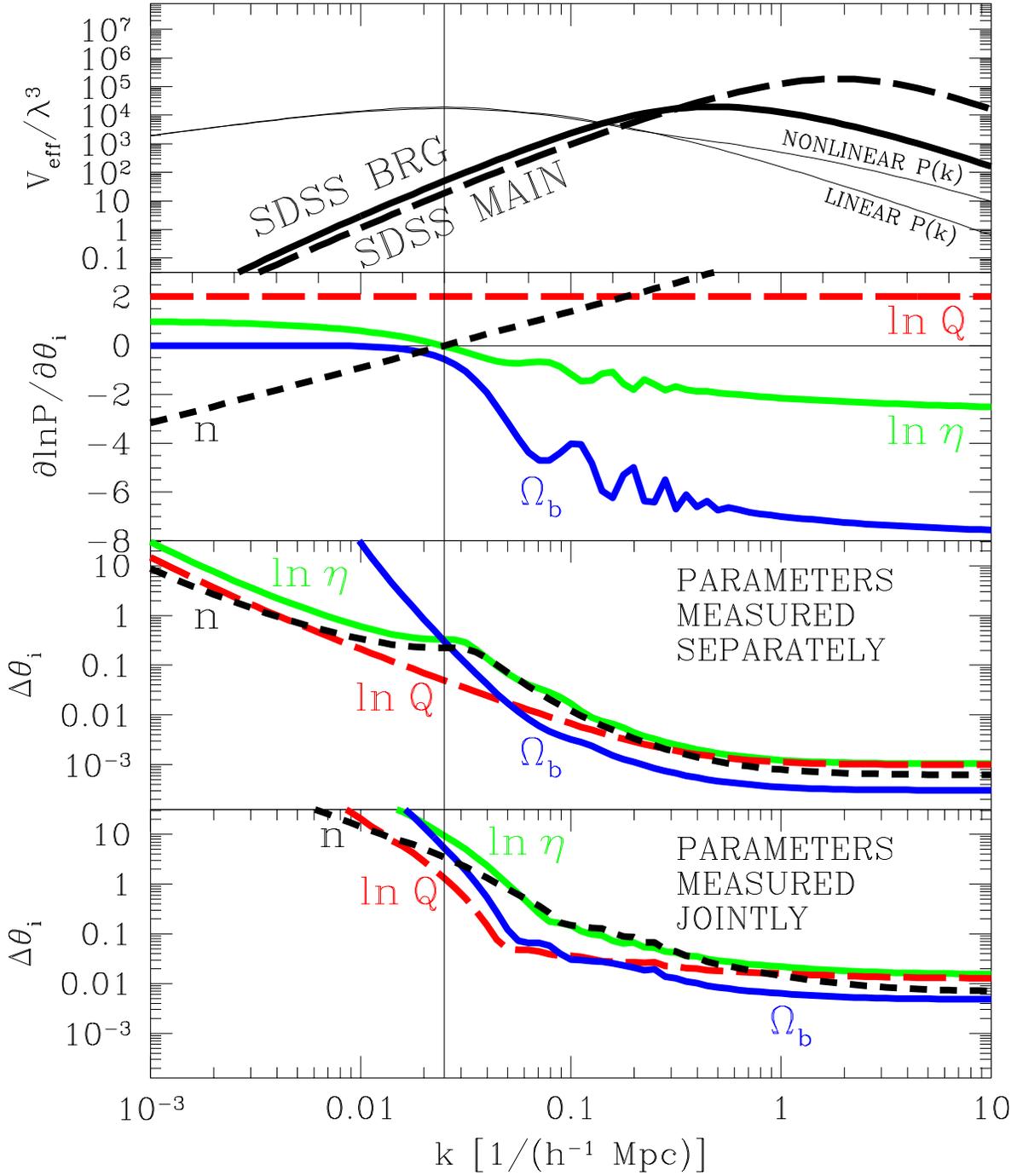}}}}
\caption{
The top panel shows the weight functions $w(k)$ for main and BRG
samples of the SDSS, together with our fiducial linear and nonlinear 
CDM power spectra in $(h^{-1}\Mpc)^3$ units.
The second panel shows the logarithmic derivatives of the linear power 
spectrum with respect to its amplitude, horizontal location, slope
and baryon content. The third panel shows the accuracy with
which these parameters can be measured using information on 
wavenumbers up to $k=\kmax$ when the other parameters are already known,
and the bottom panel shows the corresponding accuracies when 
all four parameters must be determined simultaneously.
The vertical line indicates $k_*$, the location where $P(k)$ peaks.
}
\end{figure}

\begin{figure}[phbt]
\centerline{{\vbox{\epsfxsize=20cm\epsfbox{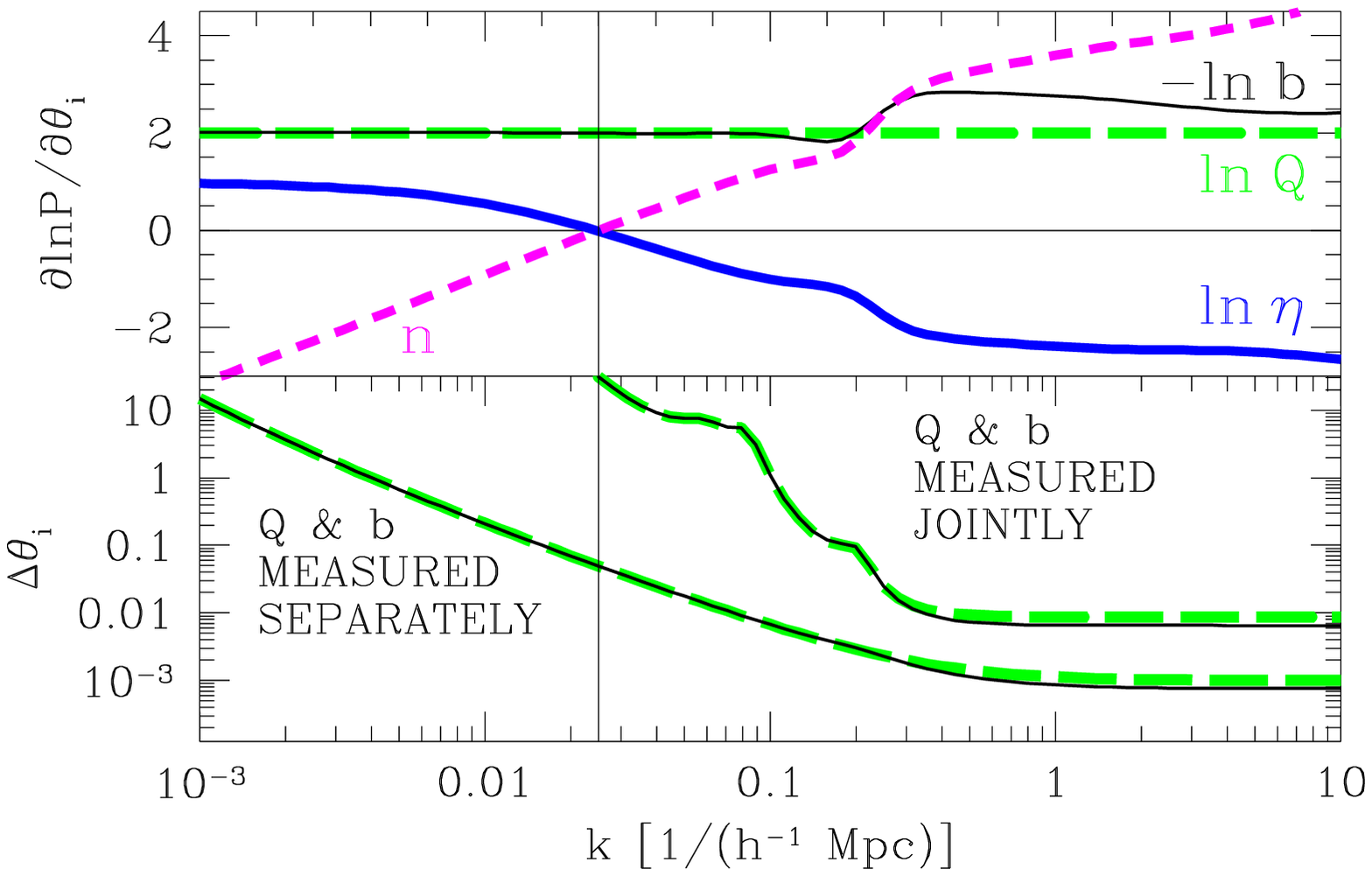}}}}
\bigskip
\caption{
Similar to Figure 1, but using nonlinear power 
spectra. In the bottom plot, all parameters except 
$Q$ and $b$ are assumed to be independently known.
}
\end{figure}



\begin{references} 



       
\bibitem{Jungman96a}
\rf G. Jungman, M. Kamionkowski, A. Kosowsky, and
D. N. Spergel;1996;Phys. Rev. Lett.;76;1007

\bibitem{Jungman96b}
\rf G. Jungman, M. Kamionkowski, A. Kosowsky,  and
D. N. Spergel;1996;Phys. Rev. D;54;1332

\bibitem{parameters}
J. R. Bond, G. Efstathiou, and M. Tegmark, preprint astro-ph/9702100.

\bibitem{Zaldarriaga97}
M. Zaldarriaga, D. Spergel, and U. Seljak 1997, preprint astro-ph/9702157.

\bibitem{Strauss97}
Strauss, M. 1997, preprint astro-ph/9610033.

\bibitem{Gunn95}
J. E. Gunn \& D. H. Weinberg, in {\it Wide-Field
Spectroscopy and the Distant Universe}, ed.\ S. J. Maddox and A.
Arag\'on-Salamanca (World Scientific, Singapore, 1995).

\bibitem{karhunen}
\rf M. Tegmark, A. N. Taylor, and A. F. Heavens;1997;ApJ;480;22

\bibitem{KendallStuart69}
M. G. Kendall \& A. Stuart, The Advanced Theory of Statistics,
Volume II (Griffin, London, 1969).

\bibitem{VS96}
\rf M. S. Vogeley \& A. S. Szalay;1996;ApJ;465;43

\bibitem{Hamilton97a}
A. J. S. Hamilton 1997, preprints astro-ph/9701008
and astro-ph/9701008, MNRAS, in press.


\bibitem{FKP}
\rf H. A. Feldman, N. Kaiser, and J. A. Peacock;1994;ApJ;426;23

\bibitem{cl}
\rf M. Tegmark;1997;Phys. Rev. D;55;5895

\bibitem{cmbfast}
\rf U. Seljak \& M. Zaldarriaga;1996;ApJ;469;437

\bibitem{BE84}
\rf J. R. Bond \& G. Efstathiou;1984;ApJ;285;L45

\bibitem{HS96}
\rf W. Hu \& N. Sugiyama;1996;ApJ;471;542 

\bibitem{Goldberg97}
D. Goldberg \& M. Strauss 1997, preprint astro-ph/9707209.

\bibitem{Jain95}
\rf B. Jain, H. J. Mo \& S. D. M. White;1995;MNRAS;276;L25


\bibitem{Weinberg95}
D. H. Weinberg 1995, in ``Wide-Field Spectroscopy and the Distant Universe",
eds. Maddox \& Arag\'on-Salamanca (World Scientific, Singapore) 

\end{references}
\end{document}